\documentclass[sigconf]{acmart}

\usepackage{breakurl}
\usepackage{amsmath,amsfonts}
\usepackage{algorithmic}
\usepackage{graphicx}
\usepackage{textcomp}
\usepackage{url}
\usepackage{psfrag}
\usepackage{xcolor}
\AtBeginDocument{%
  }

\begin{document}

\setcopyright{acmlicensed}
\copyrightyear{2023}
\acmYear{2023}
\acmDOI{10.1145/3593908.3593948}

\acmConference[GMSys '23]{Workshop on Green Multimedia Systems}{June 07--10,
  2023}{Vancouver, BC, Canada}
\acmBooktitle{Workshop on Green Multimedia Systems (GMSys '23),  June 07--10, 2023, Vancouver, BC, Canada}
\acmPrice{15.00}
\acmISBN{979-8-4007-0196-2/23/06}

\title{Video Decoding Energy Reduction  Using Temporal-Domain Filtering }

\author{Christian Herglotz, Matthias Kr\"anzler, Robert Ludwig, Andr\'e Kaup}
\email{{firstname.lastname}@fau.de}
\affiliation{%
  \institution{Chair of Multimedia Communications and Signal Processing\\ Friedrich-Alexander-Universit\"at Erlangen-N\"urnberg}
  \city{91058 Erlangen}
  \country{Germany}
}

\renewcommand{\shortauthors}{Herglotz et al.}

\begin{abstract}
In this paper, we study decoding energy reduction opportunities using temporal-domain filtering and subsampling methods. In particular, we study spatiotemporal filtering using a contrast sensitivity function and temporal downscaling, i.e., frame rate reduction. We apply these concepts as a pre-filtering to the video before compression and evaluate the bitrate, the decoding energy, and the visual quality with a dedicated metric targeting temporally downscaled sequences. We find that decoding energy savings yield $35\%$ when halving the frame rate and that spatiotemporal filtering can lead to up to $5\%$ of additional savings, depending on the content. 
\end{abstract}

%

%

\maketitle

\section{Introduction}

Today, we are living in a highly connected and globalized world. Information and communication technology (ICT) has evolved rapidly from specialized high tech for experts to mainstream technology used in the everyday life of consumers. As an example, billions of people worldwide use online video services frequently to connect to other people and be informed about the latest news. In fact, it is estimated that today, more than $50\%$ of the global Internet traffic constitutes pure video data \cite{Sandvine22}, excluding data related to services that use videos to a great extent (e.g., social networks, gaming, or messaging).  

As a consequence, it was found that the energy consumption of online video services has reached a globally significant scale \cite{ShiftFull19}, with an estimated total share of $1\%$ of global greenhouse gas (GHG) emissions. In this, the most significant part is consumed by end-user devices \cite{CarbonTrust21}. Furthermore, many portable end-user devices rely on batteries, such that low-power  and energy efficient solutions for online video applications are essential development goals both in research and industry. 

Important parts of the online video processing pipeline are the video encoder and the video decoder. Concerning the latter, 
certain studies revealed that the decoder is a major reason for a high energy consumption in portable devices such as smartphones or tablet PCs \cite{Carroll13,Herglotz20}, especially if software decoding is used \cite{Khernache21,Herglotz22a}. As a consequence, various studies focused on reducing the energy consumption of video decoders \cite{Herglotz19,Mallikarachchi20,Correa18,Herglotz20b} while 
keeping the visual quality of the video. Also for the encoder, various studies focus on the energy consumption or the complexity \cite{Vanne14,Ramasubbu22a,Zhou14} in order to obtain a high compression while using little energy resources. 

In this paper, we analyze two concepts to further reduce the energy consumption of video encoding and decoding. For the first one, we are inspired by a model that represents the human visual system (HVS) describing the sensitivity of the HVS towards different visual frequencies in the spatiotemporal frequency domain. This model is based on the spatiotemporal contrast sensitivity function (STCSF) which was first introduced in \cite{Robson66}. It is an extension of the spatial contrast sensitivity function used in imaging \cite{Barten99} and describes, depending on the spatiotemporal frequency components of a visual video signal, which components of the signal are visible and which components are invisible. 
As an application for this concept, the STCSF was used to analyze whic	h parts of a typical video processing pipeline including capturing, discretization, and display can lead to visible artifacts \cite{Watson13}. In this work, we exploit the STCSF to identify and eventually remove invisible components in a video. This removal of information leads to a lower entropy of the video signal, such that lower bitrates and ultimately lower decoding complexities are expected. 

The second concept performs a common frame rate reduction technique, where a certain number of frames is averaged and then compressed. It is well known from the literature that frame rate reductions lead to significant energy savings \cite{Herglotz20,Herglotz22a}. In this work, we evaluate the potential reductions and also analyze the impact on the perceived visual quality. 


This paper is organized as follows. First, Section~\ref{sec:lit} gives a brief overview on related literature on the STCSF and the state of the art in energy efficient video compression solutions. Afterwards, Section~\ref{sec:WoV} gives a formal definition of the STCSF, explains its meaning and intricacies in detail, and introduces our proposed algorithm and its configuration parameters. Afterwards, Section~\ref{sec:eval} presents our evaluation setup and investigates the performance of the proposed prefiltering approach. Finally, Section~\ref{sec:concl} concludes this paper.

\section{Literature Review}
\label{sec:lit}
In the literature, it is well known that quantization and subsampling techniques can lead to substantial power and energy savings \cite{Herglotz19}. In terms of quantization, it is often reported that a lower number of bits leads to a lower decoder energy consumption \cite{Herglotz19}. Also in terms of spatial scaling, it was found that a lower resolution can reduce the power consumption of a smartphone \cite{Herglotz19b}. Furthermore, considering the frame rate, it is well known that lower frame rates lead to a reduced power consumption on versatile devices \cite{Herglotz20,Herglotz22a}, but the amount of savings was not assessed. In \cite{Herrou20}, it was shown that depending on the video content, optimal frame rate estimates can be obtained to optimize the rate-distortion performance, however, the impact on the energy consumption was not discussed. 

Frame rate reduction is mostly performed by frame averaging \cite{Herrou20}. The reason is that using frame averaging, a capturing process is simulated where the shutter angle is maximized. As a consequence, moving objects are blurred. It is noteworthy that the high-frame-rate source sequence should also be captured at a very high shutter angle because otherwise, strong ghosting artifacts can occur due to object repetitions. Therefore, we select frame averaging in this paper. 

Concerning contrast sensitivity, various studies focused on the spatial contrast sensitivity \cite{Li09_CS,Peli01,Ginsburg03}, the temporal sensitivity \cite{Virsu82,Wooten10}, or both at the same time \cite{Robson66,Kelly77}. It was found that for both spatial and temporal frequencies, the contrast sensitivity can be described by a curve separating visible and invisible components of a visual signal. Considering the contrast sensitivity with respect to both temporal and spatial frequencies, it was found that a convex surface in 3D-space accurately describes the limits of human vision \cite{Robson66}.  

In the field of video processing, the concept of temporal contrast sensitivity was validated in \cite{Mackin17SMPTE}. Furthermore, contrast sensitivity was applied for quality assessment \cite{You12} and to enhance video encoding \cite{Bandoh08}. In \cite{Mackin20}, the spatiotemporal envelope of the HVS was investigated using the visibility of temporal aliasing artifacts. Furthermore, the STCSF was used to analyze the effects of capturing and displaying a video \cite{Watson13}. 



\section{Temporal Pre-filtering}
\label{sec:WoV}
\subsection{Video as a Spatiotemporal Signal}
\begin{figure}
\psfrag{w}[c][c]{ ${x_\mathrm{hor}}$}%
\psfrag{h}[r][r]{ ${x_\mathrm{ver}}$}%
\psfrag{t}[c][c]{ $t$}%
\includegraphics[width=0.4\textwidth]{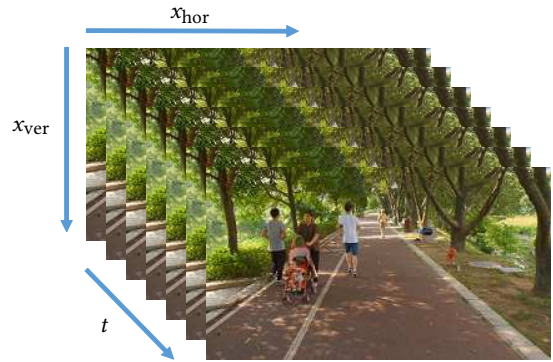}
\vspace{-0.4cm}
\caption{Definition of the visual video signal $s[\boldsymbol{x},t]$.   } 
\label{fig:signal_def}
\vspace{-0.5cm}
\end{figure}

For formalization of our methods, we define the three-dimensional, discrete video signal $s[\boldsymbol{x},t]$, where $s$ is the greyscale luminance value of the YCbCr signal, $\boldsymbol{x} = \{x_\mathrm{ver},x_\mathrm{hor}\}$ the two-dimensional spatial pixel position, and $t$ the time index (see Fig.~\ref{fig:signal_def}). 
Using this notation, we can transform the signal to the frequency domain as $S[\boldsymbol{u},w]$, where $\boldsymbol{u}$ corresponds to the horizontal and vertical spatial frequencies in the unit cycles per pixel [cpp] and $w$ to the temporal frequency in the unit frames per second [fps]. For simplification, it is common to only consider a single spatial frequency $u$ because of rotational invariance \cite{Robson66}, which means that an oscillation in any spatial direction can be transformed to a single dimension by rotation. This leads to the simplified definition of the signal $s[x,t]$ and $S[u,w]$.    

In the next step, we need to transform the visual signal, which is given in the YCbCr space, into the contrast domain. For this, we follow a concept presented in \cite{Watson05}. We thus define the contrast $c$ of a visual signal by the luminance variation divided by the mean luminance as
\begin{equation}
c = \frac{s_\mathrm{max}-s_\mathrm{min}}{s_\mathrm{mean}},
\label{eq:contr}
\end{equation}
where, to avoid conversion into physical values, we directly use the luminance component from the YCbCr signal. Hence, $s_\mathrm{max}$ and $s_\mathrm{min}$ are the maximum and the minimum luminance of the video signal, respectively. We set the mean luminance to the mean luminance of the complete video. 

The contrast is a global property of a video such that it cannot be defined for a single pixel position $[x,t]$. Therefore, we define the contrast in the frequency domain as 
\begin{equation}
C[u,w] = \frac{\left|S[u,w]\right|}{s_\mathrm{mean}},
\label{eq:ContrInFreqDomain}
\end{equation}
where the magnitude of a frequency component $\left|S[u,w]\right|$ corresponds to the amplitude of the waveform at the spatiotemporal frequency $[u,w]$. As such, the amplitude represents the luminance variation of a certain spatiotemporal frequency $[u,w]$ in the video signal $s[x,t]$. 

\subsection{Spatiotemporal Contrast Sensitiviy Function}

\begin{figure}
\psfrag{008}[tl][tl]{ $f_\mathrm{spat}$}%
\psfrag{009}[tr][tr]{ $f_\mathrm{temp}$}%
\psfrag{010}[bc][bc]{ Contrast Sensitivity  $\gamma(f_\mathrm{st})$}%
\psfrag{000}[ct][ct]{ $1$}%
\psfrag{001}[ct][ct]{ $10$}%
\psfrag{002}[rc][rc]{ $1$}%
\psfrag{003}[rc][rc]{ $10$}%
\psfrag{004}[cr][cr]{ $0$}%
\psfrag{005}[cr][cr]{ $100$}%
\psfrag{006}[cr][cr]{ $200$}%
\psfrag{007}[cr][cr]{ $300$}%
\includegraphics[width=0.4\textwidth]{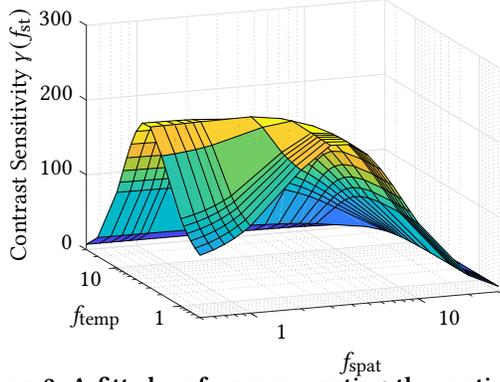}
\vspace{-0.4cm}
\caption{A fitted surface representing the spatiotemporal contrast sensitivity using Eq.~\eqref{eq:STCSF_sech} from \cite{Watson05}. The color indicates the magnitude of the contrast sensitivity.  } 
\label{fig:STCSF}
\end{figure}

Using the spatiotemporal contrast sensitivity function (STCSF) as defined in \cite{Robson66,Watson13}, our goal is now to detect frequency components that are invisible to the HVS and remove them prior to compression. 
For this, we consider the STCSF as defined on the 2D $\{u,w\}$-space. An example for such a STCSF, based on \cite{Robson66,Watson05}, is visualized in Fig.~\ref{fig:STCSF}. 
The surface in the 3D space illustrates the contrast sensitivity (vertical axis) depending on the spatial frequency and the temporal frequency (horizontal axes). 
The contrast sensitivity is defined as the reciprocal of the minimum visible contrast using the contrast as defined in Eq.~\eqref{eq:ContrInFreqDomain}. 
Note that this sensitivity is a highly simplified representation of the HVS's limits. For example, the impact of eye movement, e.g., when tracking objects in the video, is neglected. Still, as was also done in the literature \cite{Watson13}, we take the contrast sensitivity as a baseline for our work. 

Unfortunately, in the literature, we were not able to find a closed-form solution for the STCSF. Thus, as a first approximation, we propose to construct the STCSF as follows. First, we  consider the contrast sensitivity experiments reported in \cite{Robson66}, which were later discussed in detail in \cite{Watson13}. Here, it was shown that the values of the experimental cut-off frequencies, i.e. the maximum frequencies that are visible, yield approximately $f_\mathrm{spat} = 32\,$cpd for the spatial and $f_\mathrm{temp}=32\,$Hz for the temporal frequency, respectively, where cpd corresponds to the unit cycles per degree. Hence, neglecting the units, the values of the frequencies are very close. Furthermore, for lower spatial and temporal frequencies, as it is also visible in Fig.~\ref{fig:STCSF}, the surface of the STCSF shows that it is approximately rotationally invariant with respect to the spatiotemporal plane (again neglecting the units). Thus, we define the general spatiotemporal frequency 
\begin{equation}
f_\mathrm{st} = \sqrt{f_\mathrm{hor}^2 + f_\mathrm{ver}^2 + f_\mathrm{temp}^2}
\label{eq:stFreq}
\end{equation}
as the Euclidean norm of the horizontal as well as the vertical frequency $f_\mathrm{hor}$ and $f_\mathrm{ver}$, respectively, and the temporal frequency $f_\mathrm{temp}$. 
Note that $f_\mathrm{st}$ does not have a physical unit and is only meaningful when using spatial frequencies in terms of cpd and temporal frequencies in terms of Hz. 

In the next step, due to the rotational invariance property, we adopt a fit for the contrast sensitivity function defined for the spatial frequency that was proposed in \cite{Watson05} and reads 
\begin{equation}
\gamma(f_\mathrm{st}) = g\left(\mathrm{sech}\left(\left(\frac{f_\mathrm{st}}{f_0}\right)^p\right) - \alpha\cdot\mathrm{sech}\left(\frac{f_\mathrm{st}}{f_1}\right)\right). 
\label{eq:STCSF_sech}
\end{equation}
In \cite{Watson05}, it was reported that this function has the lowest root-mean square error with respect to all tested functions. The operator sech() is the hyperbolic secant, $f_0$ and $f_1$ are high- and low-frequency scales, $p$ an exponent for the high-frequency part, $\alpha$ is an attenuation factor at low frequencies, and $g$ the gain which scales to the reported contrast sensitivities in \cite{Robson66}. 
The spatiotemporal frequency $f_\mathrm{st}$ is used as the argument. We adopt the parameter values proposed in \cite{Watson05}, see Table~\ref{tab:HPmHparams}. 
\begin{table}[t]
\caption{Fitted parameter values for the STCSF in the HPmH format taken from \cite{Watson05}. }
\centering
\begin{tabular}{r|c|c|c|c|c}
\hline
Parameter & $f_0$ & $f_1$ & $\alpha$ & $p$ & $g$\\
\hline
Value & $4.1726$ & $1.3625$ & $0.8493$ & $0.7786$ & $373.08$ \\
 \hline
 \end{tabular}
\label{tab:HPmHparams}
\end{table}
With this representation, we have an analytic description of the contrast sensitivity that we exploit to identify invisible frequency components in the video signal. As this approach is an approximation, we believe that more accurate approaches might improve the results. As a first step in this direction, we test different scaling factors in the evaluation (Section~\ref{sec:eval}) and show how the compression performance changes when using different specifications of the STCSF. 

To this end, we have to convert video-domain spatial and temporal frequencies $(u,w)$ to the physical domain. Due to the usage of the FFT, the temporal frequency is calculated by 
\begin{equation}
f_\mathrm{temp} = \frac{w\cdot f_\mathrm{frame}}{N_\mathrm{temp}},
\end{equation}
where $w$ is the temporal frequency index, $f_\mathrm{frame}$ the frame rate in fps, and $N_\mathrm{temp}$ the length of the FFT in the temporal domain. 

Concerning the spatial domain, we have to convert the pixel position given in the $\boldsymbol{u}$-domain to the angular domain. To this end, we adopt the so-called Designed Viewing Distance (DVD) as defined in \cite{ITU500}, which assumes that the distance between two pixels corresponds to one arcminute angular distance on the retina of the eye. With this representation, we can avoid the use of metric units, which are otherwise needed in term of the pixel distance and the viewer's distance to the screen. This leads to the conversion factor $\mathit{\Gamma} = 60\,\frac{\mathrm{cpd}}{\mathrm{pixel}}$. Consequently, the spatial frequencies can be calculated by 
\begin{equation}
f_\mathrm{hor} = \frac{u_\mathrm{hor}\cdot \mathit{\Gamma}}{N_\mathrm{hor}}
\label{eq:freqHor}
\end{equation}
and
\begin{equation}
f_\mathrm{ver} = \frac{u_\mathrm{ver}\cdot \mathit{\Gamma}}{N_\mathrm{ver}}, 
\label{eq:freqVer}
\end{equation}
where $\{u_\mathrm{ver}, u_\mathrm{hor}\} = \boldsymbol{u}$ and $N_\mathrm{hor}$, $N_\mathrm{ver}$ are the horizontal and vertical length of the FFT, respectively.





\subsection{Signal Pruning}
\begin{figure}
\psfrag{F}[c][c]{FFT}%
\psfrag{I}[c][c]{IFFT}%
\psfrag{E}[c][c]{ \color[rgb]{1,1,1}{ENC}}%
\psfrag{>}[c][c]{ $>$}%
\psfrag{C}[c][c]{ $\beta\cdot \gamma([u,w])$}%
\psfrag{G}[c][c]{ $\downarrow$}%
\includegraphics[width=0.45\textwidth]{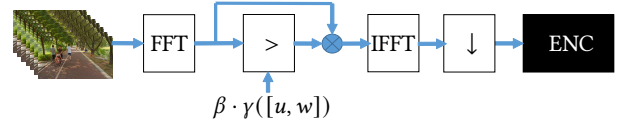}
\vspace{-0.4cm}
\caption{Workflow for removing invisible frequency components. The FFT is performed on all three dimensions.  } 
\label{fig:freq_red}
\end{figure}

Our proposed method to remove invisible frequency components is illustrated in Fig.~\ref{fig:freq_red}. 
First, we perform a three-dimensional fast Fourier transform (FFT) on the luminance part of the full input video $s[\boldsymbol{x},t]$. Afterwards, we compare the magnitude of each frequency component $\left|S[\boldsymbol{u},w]\right|$ with the minimum visible contrast $\beta\cdot \gamma(f_\mathrm{f_\mathrm{st}}) = \beta\cdot \gamma([\boldsymbol{u},w])$ and obtain the binary mask
 \begin{equation}
 M[\boldsymbol{u},w] = \left|S[\boldsymbol {u},w]\right| > \beta\cdot \gamma([\boldsymbol{u},w]), 
 \end{equation}
which can obtain a value of either one or zero. Furthermore, we define the scaling factor $\beta$ to test different scales of the STCSF. 
Afterwards, we multiply the mask with the transformed video signal entry-wise as $M[\boldsymbol{u},w]\cdot S[\boldsymbol{u},w]$  to remove the invisible frequency components. The resulting signal is inversely transformed (IFFT). After the inverse transform, we perform the temporal downscaling by frame averaging, where we choose integer downscaling factors of two and four. Finally, the resulting video is compressed by a standard video encoder. 


\section{Evaluation}
\label{sec:eval}

We use the $22$ sequences from the BVI-HFR dataset \cite{Mackin18} which are provided at HD resolution. In the temporal domain, we select $512$ frames at the original frame rate of $120\,$fps. We downsample the videos spatially to a resolution of $910\times 512$ pixels, which keeps the aspect ratio and simplifies the application of the FFT due to reduced memory requirements (FFT size of $1024\times 512\times 512$). Due to the restriction that the FFT is performed on signals with a length that is a power of two, without spatial downsampling we would have to use a FFT size of $4096\times 2048\times 512$ to allow proper transform of the video with pixel height $1080$, which is a sixteen-fold increase in memory and complexity. 
The horizontal spatial dimension is padded with zeros. 

For encoding, we use the x265 encoder \cite{x265} at medium preset with the standard constant rate factors (crf) of $18, 23, 28, 33$, and $38$. We encode the sequences at all frame rates and with the scaling factors $\beta\in\{0, 0.01, 0.05, 0.2\}$, where $\beta=0$ corresponds to no filtering.  

Concerning quality evaluation, we select a dedicated quality metric targeting temporally downscaled videos that is called Space-Time Generalized Entropic Differences (ST-GREED) \cite{Madhusudana21}. It uses statistics on spatial and temporal bandpass coefficients to come up with a quality estimate using a learned regressor. It was explicitly trained on sequences at varying frame rates. Note that a lower GREED score reflects a higher visual quality.  We refrain from using classic metrics such as PSNR or VMAF because they do not consider temporal phenomena.  

 Concerning the decoding energy evaluation, we perform energy measurements for OpenHEVC decoding \cite{openHEVC} on an Intel Core i5-4670 CPU with the help of running average power limit (RAPL) \cite{David10}. To ensure reliable measurements, we measure the decoding energy of each video bit stream multiple times until statistical validity is reached as explained in \cite{Kraenzler22}. 

We evaluate the performance as follows. First, we inspect rate-distortion curves as well as decoding-energy-distortion curves. Afterwards, we evaluate the compression performance in terms of the Bj{\o}ntegaard-Delta \cite{Bjonte01}.

\subsection{Performance Curves}
\label{secsec:RD}

\begin{figure*}
\psfrag{027}[tc][tc]{ Bitrate in Mbps}%
\psfrag{028}[bc][bc]{ GREED}%
\psfrag{030}[tc][tc]{ Bitrate in Mbps}%
\psfrag{031}[bc][bc]{ GREED}%
\psfrag{033}[tc][tc]{ Bitrate in Mbps}%
\psfrag{034}[bc][bc]{ GREED}%
\psfrag{026}[bc][bc]{ Water Splashing}%
\psfrag{029}[bc][bc]{ Guitar Focus}%
\psfrag{032}[bc][bc]{ Bobblehead}%
\psfrag{000}[ct][ct]{ $1$}%
\psfrag{001}[ct][ct]{ $10$}%
\psfrag{002}[ct][ct]{ $15$}%
\psfrag{003}[ct][ct]{ $20$}%
\psfrag{008}[ct][ct]{ $1$}%
\psfrag{009}[ct][ct]{ $14$}%
\psfrag{010}[ct][ct]{ $15$}%
\psfrag{016}[ct][ct]{ $0.1$}%
\psfrag{017}[ct][ct]{ $1$}%
\psfrag{018}[ct][ct]{ $15$}%
\psfrag{019}[ct][ct]{ $20$}%
\psfrag{004}[rc][rc]{ $25$}%
\psfrag{005}[rc][rc]{ $30$}%
\psfrag{006}[rc][rc]{ $35$}%
\psfrag{007}[rc][rc]{ $40$}%
\psfrag{011}[rc][rc]{ $16$}%
\psfrag{012}[rc][rc]{ $17$}%
\psfrag{013}[rc][rc]{ $18$}%
\psfrag{014}[rc][rc]{ $19$}%
\psfrag{015}[rc][rc]{ $20$}%
\psfrag{020}[rc][rc]{ $25$}%
\psfrag{021}[rc][rc]{ $30$}%
\psfrag{022}[rc][rc]{ $35$}%
\psfrag{023}[rc][rc]{ $40$}%
\psfrag{024}[rc][rc]{ $45$}%
\psfrag{025}[rc][rc]{ $50$}%
\psfrag{data1aa}[l][l]{\small $120\,$fps}%
\psfrag{data2}[l][l]{\small $60\,$fps}%
\psfrag{data3}[l][l]{\small $30\,$fps}%
\psfrag{data4}[l][l]{\small $\beta=0$}%
\psfrag{data5}[l][l]{\small $\beta=0.01$}%
\psfrag{data6}[l][l]{\small $\beta=0.05$}%
\psfrag{data7}[l][l]{\small $\beta=0.2$}%
\includegraphics[width=.97\textwidth]{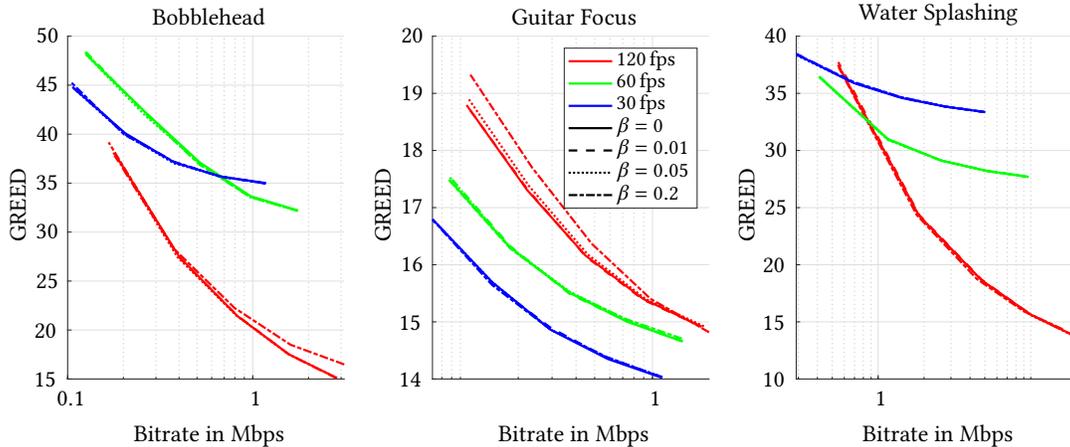}
\vspace{-0.4cm}
\caption{Visual quality in terms of GREED with respect to the bitrate for three sequences with different spatiotemporal characteristics.  } 
\label{fig:RD}
\end{figure*}

First, we evaluate filtering and temporal downscaling by rate-distortion (RD) curves illustrated in Fig.~\ref{fig:RD}. The figure shows RD curves for three selected sequences, namely Bobblehead, Guitar Focus, and Water Splashing. 

\subsubsection{Impact of Temporal Downscaling}
Concerning the impact of the frame rate (red$\,\hat =120\,$fps, green$\,\hat =60\,$fps, blue$\,\hat =30\,$fps), we observe that the RD-performance highly depends on the content of the sequence. First, for the Bobblehead sequence (left), which includes highly structured temporal and spatial frequencies (a rotating roulette), temporal downscaling leads to strong quality degradations (the curves for $60\,$fps and $30\,$fps are located above the curve for $120\,$fps). For this sequence, temporal downscaling only leads to a higher RD-performance at very low qualities (above a GREED score of $40$). 

Second, for the Guitar Focus sequence (center  of Fig.~\ref{fig:RD}), temporal downscaling always leads to a better RD performance (the $30\,$fps curve is the lowest). The Guitar Focus sequence is captured with a static camera and shows a static guitar with only slight hand and string movement. Hence, temporal downscaling leads to minor visual artifacts such that it is beneficial for the RD performance. Interestingly, the quality returned by the GREED score even increases for a lower frame rate, which is unexpected. The reason is that rate control in x265 allocates more bits for each frame at lower frame rates, because fewer frames have to be transmitted per second. Consequently, the quality for the reduced number of frames is increased significantly, which is sufficient to outweigh the quality loss of temporal downscaling. 

Third, we show results for the Water Splashing sequence (Fig.~\ref{fig:RD}, right), which includes highly random  spatial and temporal frequencies. In this case, the original sequence at $120\,$fps can reach highest qualities, similar to the Bobblehead sequence. In contrast, temporal downscaling leads to a better RD performance at a lower GREED score of roughly $30$. 

Summarizing, we find that the impact of temporal downscaling on the RD performance highly depends on the content of the sequence. The method is highly effective when  scenes are static and most ineffective when there are highly structured spatial and temporal frequencies. 

\subsubsection{Impact of Spatiotemporal Filtering}

The impact of filtering is visualized by the line style (solid for no filtering $\beta=0$, dashed for $\beta=0.01$, dotted for $\beta=0.05$, and dashed-dotted for $\beta=0.2$). We observe that the impact on the RD performance is much smaller than for temporal downscaling, which can be expected because the number of samples to be encoded is not changed. Similar to downscaling, we can see that the impact of filtering highly depends on the sequence. For Water Splashing at $120\,$fps and high visual qualities (red lines), filtering increases the RD performance slightly. We also observe improvements for the Guitar Focus sequence at $30\,$fps between GREED scores of $15$ and $16$. In some cases, however, filtering leads to a lower RD performance (e.g., Bobblehead and Guitar Focus at $120\,$fps).

\subsubsection{Decoding Energy Savings}
\begin{figure*}
\psfrag{036}[tc][tc]{ Decoding Energy in J}%
\psfrag{037}[bc][bc]{ GREED}%
\psfrag{039}[tc][tc]{ Decoding Energy in J}%
\psfrag{040}[bc][bc]{ GREED}%
\psfrag{042}[tc][tc]{ Decoding Energy in J}%
\psfrag{043}[bc][bc]{ GREED}%
\psfrag{025}[bc][bc]{ Pond}%
\psfrag{028}[bc][bc]{ Bouncyball}%
\psfrag{031}[bc][bc]{ Bobblehead}%
\psfrag{035}[bc][bc]{ Water Splashing}%
\psfrag{038}[bc][bc]{ Guitar Focus}%
\psfrag{041}[bc][bc]{ Bobblehead}%
\psfrag{000}[ct][ct]{ $0$}%
\psfrag{001}[ct][ct]{ $20$}%
\psfrag{002}[ct][ct]{ $40$}%
\psfrag{003}[ct][ct]{ $60$}%
\psfrag{004}[ct][ct]{ $80$}%
\psfrag{012}[ct][ct]{ $0$}%
\psfrag{013}[ct][ct]{ $5$}%
\psfrag{014}[ct][ct]{ $10$}%
\psfrag{015}[ct][ct]{ $15$}%
\psfrag{023}[ct][ct]{ $0$}%
\psfrag{024}[ct][ct]{ $10$}%
\psfrag{025}[ct][ct]{ $20$}%
\psfrag{026}[ct][ct]{ $30$}%
\psfrag{005}[rc][rc]{ $10$}%
\psfrag{006}[rc][rc]{ $15$}%
\psfrag{007}[rc][rc]{ $20$}%
\psfrag{008}[rc][rc]{ $25$}%
\psfrag{009}[rc][rc]{ $30$}%
\psfrag{010}[rc][rc]{ $35$}%
\psfrag{011}[rc][rc]{ $40$}%
\psfrag{016}[rc][rc]{ $14$}%
\psfrag{017}[rc][rc]{ $15$}%
\psfrag{018}[rc][rc]{ $16$}%
\psfrag{019}[rc][rc]{ $17$}%
\psfrag{020}[rc][rc]{ $18$}%
\psfrag{021}[rc][rc]{ $19$}%
\psfrag{022}[rc][rc]{ $20$}%
\psfrag{027}[rc][rc]{ $15$}%
\psfrag{028}[rc][rc]{ $20$}%
\psfrag{029}[rc][rc]{ $25$}%
\psfrag{030}[rc][rc]{ $30$}%
\psfrag{031}[rc][rc]{ $35$}%
\psfrag{032}[rc][rc]{ $40$}%
\psfrag{033}[rc][rc]{ $45$}%
\psfrag{034}[rc][rc]{ $50$}%
\psfrag{data1aa}[l][l]{\small $120\,$fps}%
\psfrag{data2}[l][l]{\small $60\,$fps}%
\psfrag{data3}[l][l]{\small $30\,$fps}%
\psfrag{data4}[l][l]{\small $\beta=0$}%
\psfrag{data5}[l][l]{\small $\beta=0.01$}%
\psfrag{data6}[l][l]{\small $\beta=0.05$}%
\psfrag{data7}[l][l]{\small $\beta=0.2$}%
\includegraphics[width=.97\textwidth]{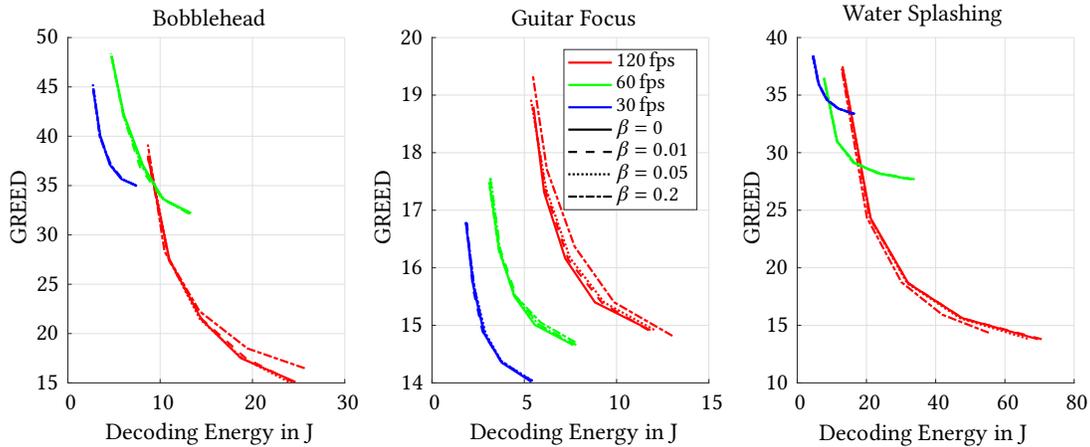}
\vspace{-0.4cm}
\caption{Visual quality in terms of GREED with respect to the decoding energy for three sequences with different spatiotemporal characteristics.  } 
\label{fig:DED}
\end{figure*}

The decoding energy versus the quality is plot in Fig,~\ref{fig:DED} for the same sequences. 
For the Bobblehead and the Water Splashing sequence (left and right), we find that at low qualities (GREED score above 35), temporal downscaling leads to a lower energy consumption. For the Guitar Focus sequence, similar to the RD performance, temporal downscaling is always the better choice. 

With respect to filtering, we again find that the impact is much lower than the impact of temporal filtering. Some decoding energy savings are obtained when also RD savings are observed (e.g., the Guitar Focus sequence at $30\,$fps and a GREED score of $15.5$, Water Splashing at $120\,$fps and GREED scores below $25$).

\subsection{Average Savings}
To assess the amount of savings, we calculate average bitrate savings and average decoding energy savings over a constant visual quality using the B{\o}ntegaard-Delta \cite{Bjonte01}. We use Akima interpolation as suggested in \cite{Herglotz22b} and select the GREED score as the quality metric. To calculate decoding energy savings, we replace the  bitrate with the decoding energy in the BD calculus as performed in \cite{Kraenzler22}. The reference for BD calculations is the compression of the unfiltered sequence at $120\,$fps. 

It is important to mention that the average savings are calculated over the overlapping range of GREED scores. Consequently, the BD values are valid for different quality ranges. For example, in the case of Bobblehead, the overlap at different frame rates ranges from roughly $35$ to $40$, and for Guitar Focus from roughly $15$ to $17$. 

\subsubsection{Temporal Downscaling}
Table~\ref{tab:BD} lists BD-rate savings and BD-decoding energy savings for temporal downscaling to $60\,$fps and $30\,$fps. 
\begin{table}[t]
\caption{Relative bitrate savings and decoding energy savings for different frame rates in percent.  }
\vspace{-0.4cm}
\label{tab:BD}
\centering
\begin{tabular}{r||c|c|c||c|c|c}
 & \multicolumn{3}{c||}{BD-rate} & \multicolumn{3}{c}{BD-Decoding Energy}\\
 & min  & mean & max & min & mean  & max  \\ 
 \hline
 $60\,$fps & $-76.75$ & $-6.52$ & $215.55$ & $-61.84$ & $-35.01$ & $-1.13$\\
 $30\,$fps & $-92.88$ & $-21.41$ & $125.09$ & $-83.99$ & $-54.16$ & $-45.65$\\
 \end{tabular}
 \end{table}
We neglect sequences where no overlap of GREED scores occured. The table shows that on average, significant bitrate reductions as well as decoding energy reductions can be obtained. 

Concerning the rate, we find mean rate savings of more than $6\%$ for $60\,$fps and more than $20\%$ for $30\,$fps. However, the range of values shows a very high variability. While highest rate savings reach up to $93\%$ (static sequence like Guitar Focus), in some cases the rate even increases (e.g., Bobblehead). This, again, proves that temporal downscaling should only be performed for certain content. 

Regarding the decoding energy, mean savings are significantly higher ($35\%$ for $60\,$fps and $54\%$ for $30\,$fps). Also, the variability of savings is lower, but still significant (between $1\%$ and $62\%$ for $60\,$fps and between $45\%$ and $84\%$ for $30\,$fps). In general, we find that high decoding energy savings occur when we also observe high bitrate savings. 

\subsubsection{Spatiotemporal Filtering}
Concerning the filtering, we observe more variability. We calculate BD values of the filtered and compressed sequences with respect to the unfiltered sequences at the same frame rate. This highlights the pure impact of filtering, independent from frame rate changes. Analyzing these values, we find that in $40\%$ and $42\%$ of cases (i.e. sequences and scaling factors $\beta$), the bitrate and the decoding energy, respectively, is reduced. Corresponding maximum rate and energy savings yield $7.7\%$ and $5.6\%$, respectively, where both are observed for the Sparkler sequence. 

For the Water Splashing sequence (right of Figs.~\ref{fig:RD} and \ref{fig:DED}), maximum savings yield $2\%$ bitrate and $4.5\%$ decoding energy savings, respectively. Averaging over all sequences, mean bitrate and energy savings are marginal for all scaling factors (absolute mean savings smaller than $0.5\%$). Future work could further investigate this behavior, identify relations between the content and actual savings, and develop a content-adaptive filtering solution.

\section{Conclusion}
\label{sec:concl}
This paper analyzed a spatiotemporal filtering technique with temporal downscaling on visual quality, bitrate, and decoding energy savings. Our evaluations indicate that frame rate reduction is a powerful method to reduce the bitrate and the decoding energy substantially. When halving the frame rate, we observe mean bitrate savings of $6.5\%$ and mean decoding energy savings of $35\%$. Concerning filtering, we find that for certain video content, the bitrate and the decoding energy can be further reduced by more than $7\%$ and $5\%$, respectively. However, these savings highly depend on the content of the sequence and need further investigation. 

Future work can exploit this knowledge to generate an adaptive frame rate reduction method that, depending on the content and the target quality, decides the optimal frame rate and filtering method. In addition, the proposed global filtering method could be replaced by local filtering. Furthermore, the approach could be combined with spatial downsampling methods to obtain optimal spatiotemporal scaling. 

  \bibliographystyle{ACM-Reference-Format}
  \bibliography{literature}

%
%

\end{document}